\documentclass[onecolumn,showpacs,floatfix]{revtex4}

\usepackage{dcolumn}
\usepackage{bm}

\usepackage{amssymb,amsmath,epsfig,graphicx}

\newcommand{\vmedia}{\langle \dot x_{\text{cm}}\rangle}  

\begin{document}

\title{Transport reversal in a delayed feedback ratchet}

\author{M. Feito}
\email{feito@fis.ucm.es}
\affiliation{Departamento de F\'{\i}sica At\'omica, Molecular y
Nuclear, Universidad Complutense de Madrid, \\
Avenida Complutense s/n, 28040 Madrid, Spain.}

\author{F. J. Cao}
\email{francao@fis.ucm.es}
\affiliation{Departamento de F\'{\i}sica At\'omica, Molecular y
Nuclear, Universidad Complutense de Madrid, \\
Avenida Complutense s/n, 28040 Madrid, Spain} \affiliation{LERMA,
Observatoire de Paris, Laboratoire Associ\'e au CNRS UMR 811 2,
\\ 61, Avenue de l'Observatoire, 75014 Paris, France.}


\begin{abstract}
Feedback flashing ratchets are thermal rectifiers that use
information on the state of the system to operate the switching on and
off of a periodic potential. They can induce directed transport even
with symmetric potentials thanks to the asymmetry of the feedback
protocol. We investigate here the dynamics of a feedback flashing
ratchet when the asymmetry of the ratchet potential and of the
feedback protocol favor transport in opposite directions. The
introduction of a time delay in the control strategy allows one to
nontrivially tune the relative relevance of the competing 
asymmetries leading to an
interesting dynamics. We show that the competition between the
asymmetries leads to a current reversal for large delays. For
small ensembles of particles current reversal appears as the
consequence of the emergence of an open-loop like dynamical
regime, while for large ensembles of particles it can be
understood as a consequence of the stabilization of quasiperiodic
solutions. We also comment on the experimental feasibility of these feedback
ratchets and their potential applications.
\end{abstract}

\pacs{05.40.-a, 02.30.Yy}
\maketitle


\section{Introduction}

Brownian motors or ratchets are spatially periodic systems that
are able to induce direct transport rectifying thermal
fluctuations. Two conditions are generally sufficient for the
emergence of direct transport in these systems: breaking of
thermal equilibrium and breaking of spatial inversion symmetry
\cite{rei02}. These systems permit one to get an insight into
non-equilibrium processes and are receiving
increasing interest also due to their applications in
nanotechnology and biology~\cite{rei02,lin02, kay07}.
\par

Flashing ratchets are devices that rectify the motion of Brownian
particles by subjecting them to a spatially periodic potential
that is alternatively switched on and off. Open-loop flashing
ratchets operate without regard to the state of the system (open-loop control)
implementing a periodic or random switching to
rectify thermal fluctuations by taking advantage of the asymmetry
of the potential~\cite{bug87,ajd93,ast94}. On the contrary,
feedback ratchets (or closed-loop ratchets) use information on the
particle distribution of the system to
operate~\cite{cao04,cao07,fei07,fei07b,cra07,din05,fei06}, and the
asymmetry of the feedback control protocol is able to induce a
directed transport even for symmetric ratchet potentials. For
instance, in the so-called maximization of the center-of-mass
velocity protocol~\cite{cao04} the controller switches on the
potential only if switching on would imply a positive displacement
for the center-of-mass position (i.e., if the net force with the
potential on would be positive). Feedback flashing ratchets have
been recently suggested as a mechanism to explain the stepping
motion of the two-headed kinesin~\cite{bie07}. In another context, a feedback
scheme has been used to perform control of chaotic trajectories in 
inertia ratchets~\cite{vin07}.
\par

Feedback flashing ratchets could be experimentally implemented
monitoring the positions of a set of Brownian
particles~\cite{rou94,mar02,coh06} and subsequently using the
information gathered to decide whether to switch on or off a
ratchet potential according to a giving protocol. This
experimental design will have to deal with a finite time lag
between the collection of the information about the state of the
system and the action because of the time interval needed for the
measurement, transmission and processing of the
information~\cite{ste94,bec05}. Time delays in the feedback also
appear naturally in complex systems with self regulating
mechanisms (see~\cite{boc00,fra05b} and references therein). It is also
remarkable for the ability of controlling chaos and improving
coherence in excitable systems under delayed feedback~\cite{sch07,pra07}. The
feasibility of nanotechnological feedback flashing ratchet devices
and their performance under the presence of a time delay has been
analyzed very recently in Refs.~\cite{fei07b,cra07}. In those works, and
also in previous
ones~\cite{cao04,cao07,fei07,din05,fei06}, the two
sources of spatial asymmetry involved, namely the feedback control
and the shape of the potential, \emph{cooperate} with the aim of
maximizing the performance of the system. On the contrary, in this
paper we investigate the  effects of the
\emph{competition} between the potential asymmetry and the control asymmetry
in a delayed feedback ratchet.
\par

We have observed a rich dynamics that includes transport reversal. The
inversion of the current direction upon the variation of the 
system parameters is a well-known phenomenon in Brownian motors that can be
produced by varying the characteristics of the non-equilibrium
fluctuations~\cite{doe94,mil94} or the parameters of the time-dependent
perturbation that drives the system out of
equilibrium~\cite{bar94,dan01,ai05,chau95,bie96}. It also appears in other
ratchet-like systems, such as deterministic inertial
ratchets~\cite{mat00,bar00}. The phenomenon of current reversal has great
importance in particle separation devices~\cite{ket00}, and in biology
systems~\cite{hen97}. In our present study current reversal is
achieved just by varying the time delay of the system.
\par

We start below with the description of the collective flashing ratchet and the
delayed feedback protocol that we consider. In the next section,
Sec.~\ref{sec:results}, 
the evolution equations of the system are solved by Langevin dynamics
simulations and the rich dynamics encountered (transport reversal,
quasi-periodic modes of oscillation, multistability) is analyzed. We finally
review and further discuss in Sec.~\ref{sec:discussion} the implications of
the results.

\section{Model}
\begin{figure}
\includegraphics[scale=0.5]{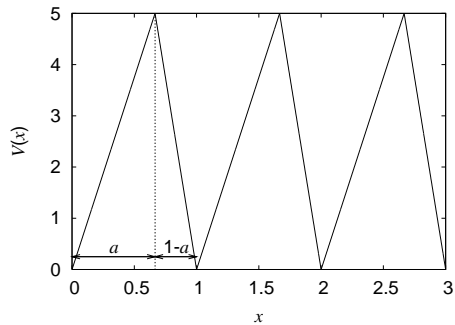}
\caption{\label{fig:potential}
Sawtooth potential [Eq.~\eqref{sawtoothpot}] of height $V_0=5k_BT$ and
asymmetry parameter $a=2/3$. Units: $L=1$ and $k_BT=1$.}
\end{figure}

The feedback ratchet that we consider consists of $N$ Brownian
particles at temperature $T$ in a periodic potential $V(x)$. The
force acting on the particles is $F(x)=-V^{\prime}(x)$, where the
prime denotes the spatial derivative. The state of this system is
described by the positions $x_i(t)$ of the particles satisfying
the overdamped Langevin equations
\begin{equation}\label{langevin}
\gamma \dot x_i(t)=\alpha(t)F(x_i(t))+\xi_i(t);\quad i=1,\dots,N,
\end{equation}
where $\gamma$ is the friction coefficient (related to the
diffusion coefficient $D$ through Einstein's relation
$D=k_BT/\gamma$), $\xi_i(t)$ are Gaussian white noises of zero
mean and variance $\langle \xi_i(t)\xi_j(t^\prime)\rangle =2\gamma
k_B T\delta_{ij}\delta(t-t^\prime)$, and $\alpha(t)$ stands for the action of
the controller. The feedback policy uses the sign of the net force
per particle,
\begin{equation}
f(t)=\frac{1}{N}\sum_{i=1}^N F(x_i(t)),
\end{equation}
as follows: The controller measures the sign of the net force and,
after a time $\tau$, switches the potential on ($\alpha=1$) if
the net force was positive or switches the potential off
($\alpha=0$) if the net force was negative. Therefore, the delayed
control protocol considered is
\begin{equation}
  \alpha(t)=
  \begin{cases}
    \Theta(f(t-\tau)) &\text{if } t\geq \tau ,\\
    0 &\text{otherwise},
  \end{cases}
\end{equation}
with $\Theta$ the Heaviside function [$\Theta (x)=1$ if $x>0$,
else $\Theta (x)=0$].
We have used a sawtooth potential of period $L$,
i.e. $V(x)=V(x+L)$, height $V_0$, and asymmetry parameter $a$:
\begin{equation}\label{sawtoothpot}
  V(x)=
  \begin{cases}
    \frac{V_0}{a}\frac{x}{L} &\text{if } 0\leq \frac{x}{L}\leq a ,\\
    V_0-\frac{V_0}{1-a}\left(\frac{x}{L}-a\right) &\text{if } a<
    \frac{x}{L}\leq 1.
  \end{cases}
\end{equation}
The height $V_0$ of the potential is the difference
between the value of the potential at the minimum and at the maximum, while
$aL$ is the distance between the minimum and the maximum consecutive
positions (Fig.~\ref{fig:potential}). Thus when $a<1/2$ both the asymmetry of
the potential and the feedback protocol favor transport in the same direction,
whereas when $a>1/2$ there is a competition between them. We consider here the
latter case.

\section{Results}\label{sec:results}

We have performed numerical simulations of the Langevin
equations~\eqref{langevin} by using an Euler-Maruyama
scheme~\cite{klo92}, which reveals different dynamics of the
collective ratchet for different ensemble sizes when the delayed
feedback is present. We distinguish between few particles
(including $N=1$ as a particular case) and many particles.
Following previous works~\cite{cao04}, we refer to the few
particle case when the average long-time limit velocity of the
center-of-mass, $\vmedia$, is greater for the non-delayed feedback
maximization protocol than for the optimal open-loop protocol, and
many particle case otherwise. Typically the frontier between these
regimes corresponds to $N=10^2-10^3$ particles for potential
heights of the order of $5k_BT$ or greater.
\par

Let us begin our analysis with the non-delayed ($\tau=0$) feedback ratchet.

\subsection{Non-delayed feedback ratchet}\label{subsec:non}
For \emph{one particle} ($N=1$) the exact expression for the
average velocity can be obtained by solving a Fokker-Planck
equation with the proper effective potential that includes the
action of the controller. This expression has been derived
in~\cite{cao04}, and is indeed valid for any asymmetry parameter
$0<a<1$. It gives a positive flux that grows as $D V_0/(k_BT L)$
for small potential heights ($V_0 \lesssim k_BT $), and tends to
the finite value $2D/(a^2L)$ for large potential heights ($V_0 \gg
k_BT $). The fact that the flux goes to a constant value for large
potential heights is a direct consequence of the overdamped nature
of the ratchet. We remark that enlarging the value of the ratio
$V_0/(k_B T)$ corresponds to effectively diminish the intensity of the white
noise that accounts for thermal fluctuations.
\par

\begin{figure}
\includegraphics[scale=0.60]{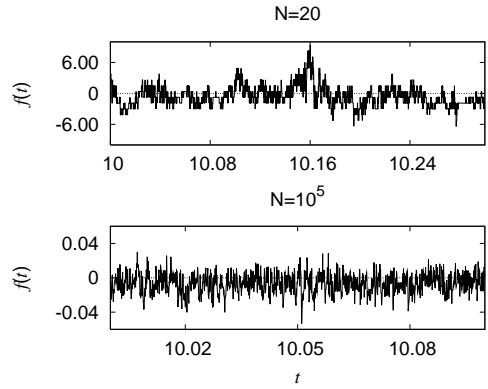}
\caption{\label{fig:evol_nodelay}
Evolution of the net force per particle for the non delayed feedback ratchet
for $N=20$ (few particles) and $N=10^5$ (many particles). The potential is
``on'' only in the time intervals such that $f(t)$ is positive. Parameters of
the potential: $V_0=5k_BT$ and $a=2/3$. Units: $L=1$, $D=1$, and $k_BT=1$.
}
\end{figure}

For the collective ratchet compounded of a \emph{few particles} an
approximation for the center-of-mass velocity can be obtained
assuming a purely stochastic behavior (see Ref.~\cite{cao04} for
details). As the magnitude of the fluctuations of the force are of
the order of the inverse of the square-root of the number of
particles the stochastic approximation predicts the $\vmedia\sim
1/\sqrt{N}$ decay observed in our simulations for any asymmetries.
In fact this qualitative behavior remains valid for any number of
particles (including \emph{many particles}) provided the asymmetry
is $a>1/2$. In this latter case the potential asymmetry acts
against the feedback protocol, which tries to favor positive
currents, and then the potential is turned on in very small
intervals of time as the controller rapidly switches it off. See
Fig.~\ref{fig:evol_nodelay}. Thus the systems dynamics is
effectively stochastic, contrary to the many particle case with
cooperating asymmetries, where switches are slower and allow the
system to have enough time to evolve in a quasideterministic
way~\cite{cao04}. Therefore when the potential asymmetry competes
against the feedback the flux decays with the number of particles
as $1/\sqrt{N}$ even for many particles, contrary to the much
slower $1/\ln N$ dependence observed when both asymmetries
cooperate~\cite{cao04}.
\par

In any case the non-delayed protocol always gives a positive flux
because it only switches on when it implies a positive
displacement of the center-of-mass position.

\subsection{Delayed feedback ratchet}
The presence of a lag time in the control can cause negative currents and
complicated dynamics that depends on the number of particles. Let us first
study the few particle case (including one particle).

\subsubsection{Few particles}

\begin{figure}
\includegraphics[scale=0.5]{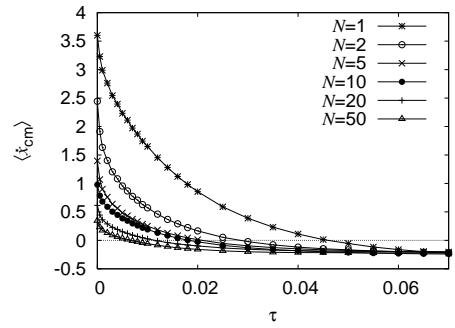}
\caption{\label{fig:flux_V0_10_N}
Center-of-mass velocity $\vmedia$ as a function of the time delay $\tau$ for
different numbers $N$ of particles under the few particle
regime. Parameters of the potential: $V_0=10k_BT$ and $a=2/3$. Units: $L=1$,
$D=1$, and $k_BT=1$.
}
\end{figure}

\begin{figure}
\includegraphics[scale=0.5]{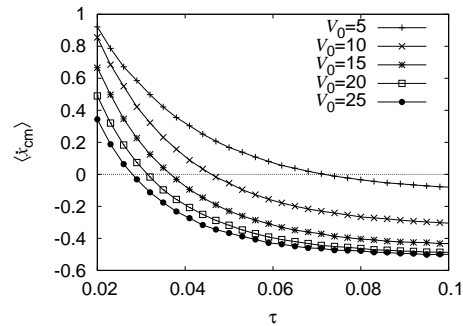}
\caption{\label{fig:flux_various_V0}
One particle velocity versus the time delay for
different heights $V_0$ of the potential and asymmetry parameter $a=2/3$.
Units: $L=1$, $D=1$, and $k_BT=1$.
}
\end{figure}

When the control protocol presents a time delay the system
performs worse because the delayed action of the controller
implies some wrong actions. Moreover, for large time delays the controller
is unable to surmount the potential shape asymmetry and eventually the net
current becomes negative. See Figs.~\ref{fig:flux_V0_10_N}
and~\ref{fig:flux_various_V0}. 

\par

For increasing time delays the correlation between the present
sign of the net force and the measured sign that the controller
actually uses decreases. Thus the controller action begins to be
uncorrelated to the present state of the system and it effectively
begins to act as an open-loop ratchet~\cite{fei07b}. In fact, for
large delays the correlation between the state of the system and
the measured retarded state is negligible and the negative flux
becomes independent of the delay,
because the ratchet is effectively open-loop controlled; see
Fig.~\ref{fig:flux_V0_10_N}. Therefore transport reversal appears
here as a consequence of the competition between the asymmetry of
the ratchet potential and the inherent asymmetry of the protocol.
We stress that the influence of the asymmetry of the feedback
protocol itself is not tuned here trivially, but changing the delay $\tau$ in
the control. Other ways of tuning the influence of the feedback
protocol could not lead to current reversal. For example, it can
be shown that a feedback protocol that switches on/off following
the maximization protocol but with a probability of error
$0<p<1/2$ does not enable negative fluxes even for asymmetries
$a>1/2$ (see Ref.~\cite{cao07}).
\par

For the delayed protocol considered the critical value of the
delay that gives zero current (thus the current is positive for
smaller delays and negative for larger delays) is related with the
characteristic time in which the information about the state of
the system is effectively lost, so the delayed maximization
protocol is not able to achieve its goal of producing a positive
current for delays larger than the critical one. Increasing the
height $V_0$ of the potential implies a faster dynamics. Therefore
the critical delay is expected to decrease with the height of the
potential, in agreement with our simulations; see
Fig.~\ref{fig:flux_various_V0}. It is important to note that this
critical delay tends to a constant \emph{nonzero} value as $V_0\to\infty$.
The reason is the same that makes the absolute value of the
flux in both closed-loop and open-loop ratchets does not grow
indefinitely as the potential goes up. These fluxes tend to a
finite value because the time spent by the particles in diffusing during the
off potential state goes to a constant in the absence of inertia.
Note also that the critical delay decreases with the number of
particles (see Fig.~\ref{fig:flux_V0_10_N}.)
\par

Let us now study the many particle case, which exhibits a
completely different dynamics.

\subsubsection{Many particles}
\begin{figure}
\includegraphics[scale=0.5]{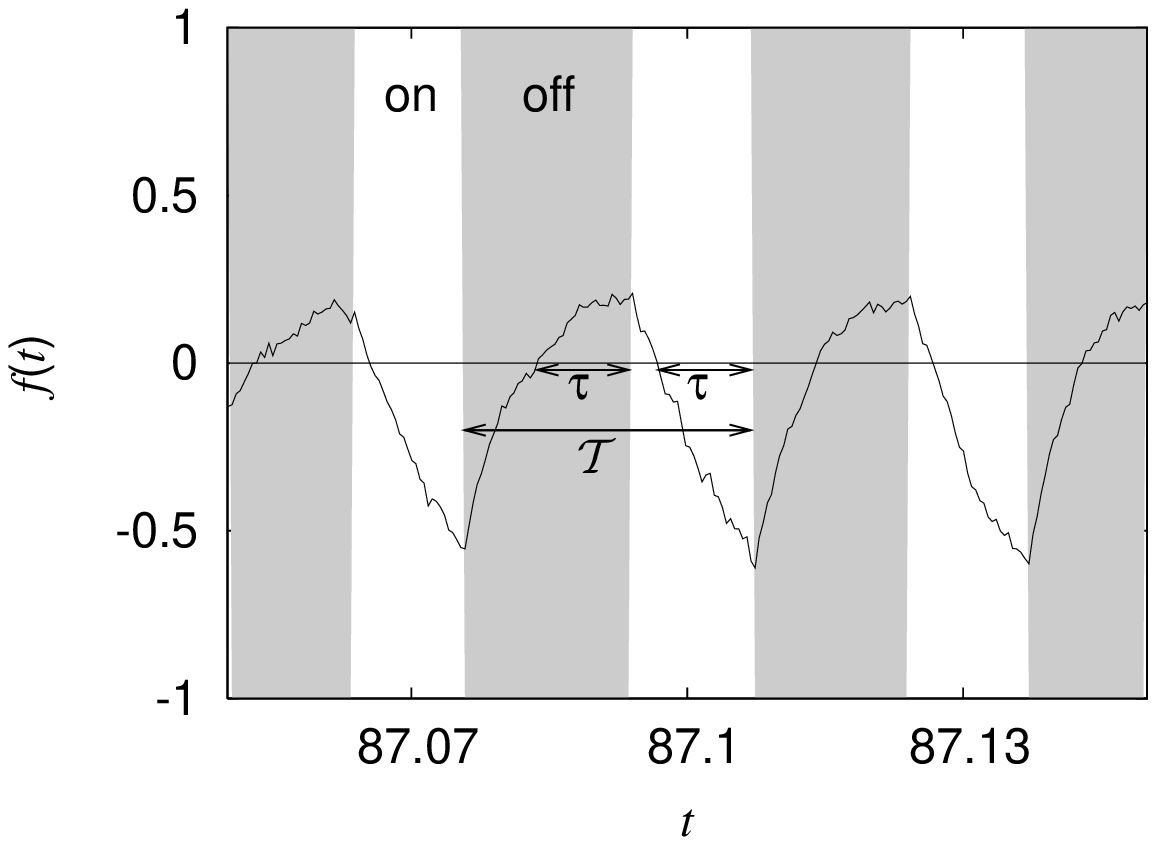}
\includegraphics[scale=0.5]{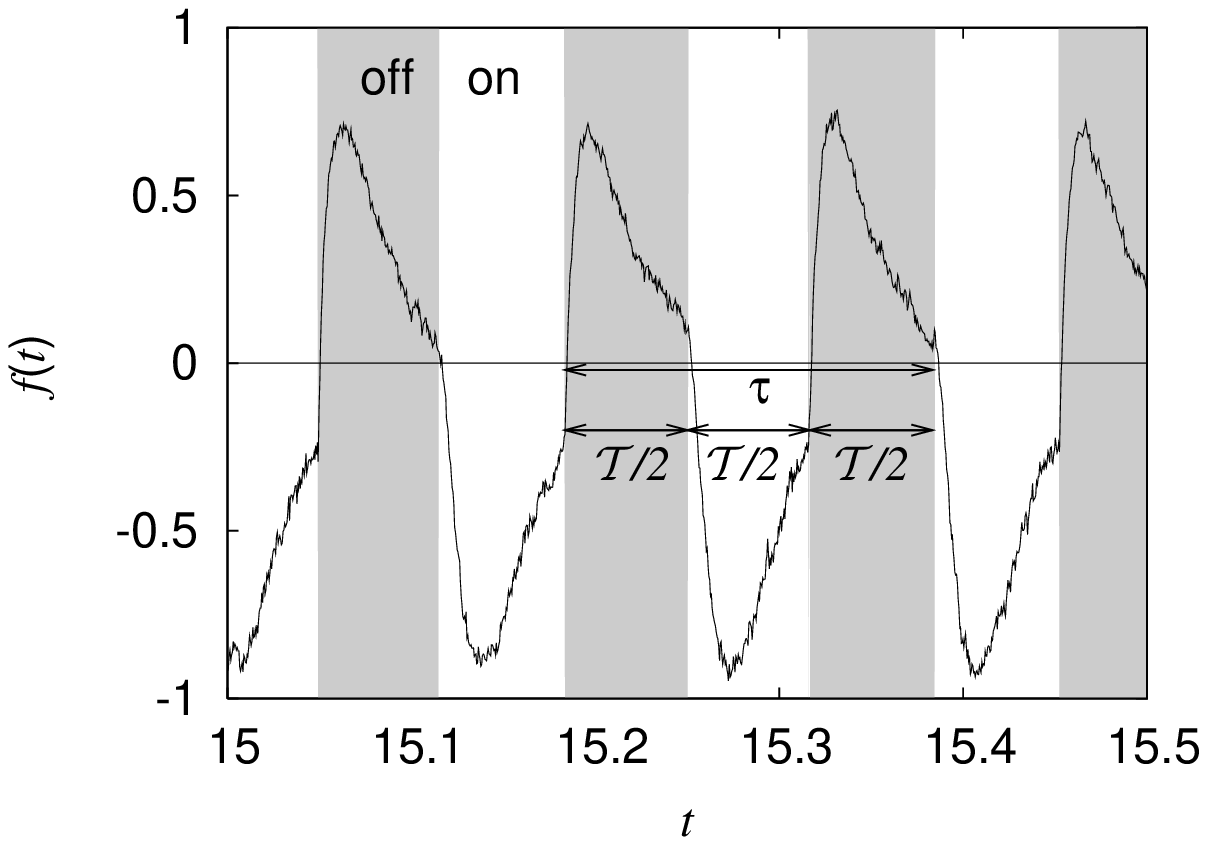}
\caption{\label{fig:evol}
Evolution of the net force per particle for time delays $\tau=0.01$ (left) and
$\tau=0.20$ (right) in the many particle case
($N=10^5$). White background regions stand for ``on''
potential and gray background regions for ``off'' potential. Parameters of the
potential: $V_0=5k_BT$ and $a=2/3$. Units: $L=1$, $D=1$, and $k_BT=1$.
}
\end{figure}

For large ensembles of particles ($N>10^2-10^3$) the flux is nearly zero in the
non-delayed protocol (see Sec.~\ref{subsec:non}), but the introduction of a
time 
delay stabilizes quasiperiodic solutions that give noticeable negative
currents. We have found that, after a transient time, the delayed control
allows the system to synchronize into a stable mode of oscillation such that
the net force per particle evolves quasi-periodically. The evolution is not
strictly periodic due to the stochastic nature of the dynamics. See
Fig.~\ref{fig:evol}. 
\par

For small delays the on and off times of the non-delayed dynamics
(Sec.~\ref{subsec:non}) are enlarged owing to the delay, and the net force
per particle evolves with a more regular pattern (Fig.~\ref{fig:evol}, left).
When the potential is switched on the net force per particle begins to
diminish (because the potential asymmetry acts against the
feedback protocol) and rapidly gets negative, but the potential still remains
`on' during a time $\tau$ after the force changed its sign. On the other hand,
when the potential is switched off the net force grows, becomes positive, and
induces an `on' switching a time $\tau$ later. The result is a quasiperiodic
dynamics with an small quasiperiod ${\cal T}>2\tau$; see Fig.~\ref{fig:evol}
(left). We highlight that these types of solution are only observed for
asymmetries $a>1/2$, as they are consequences of the competing asymmetry of the
potential; they do not appear for asymmetries $a<1/2$ that support positive
transport and exhibit a different behavior related with the enlargement of
the tails of the net force per particle~\cite{fei07b,cra07}.
\par

\begin{figure}
\includegraphics[scale=0.5]{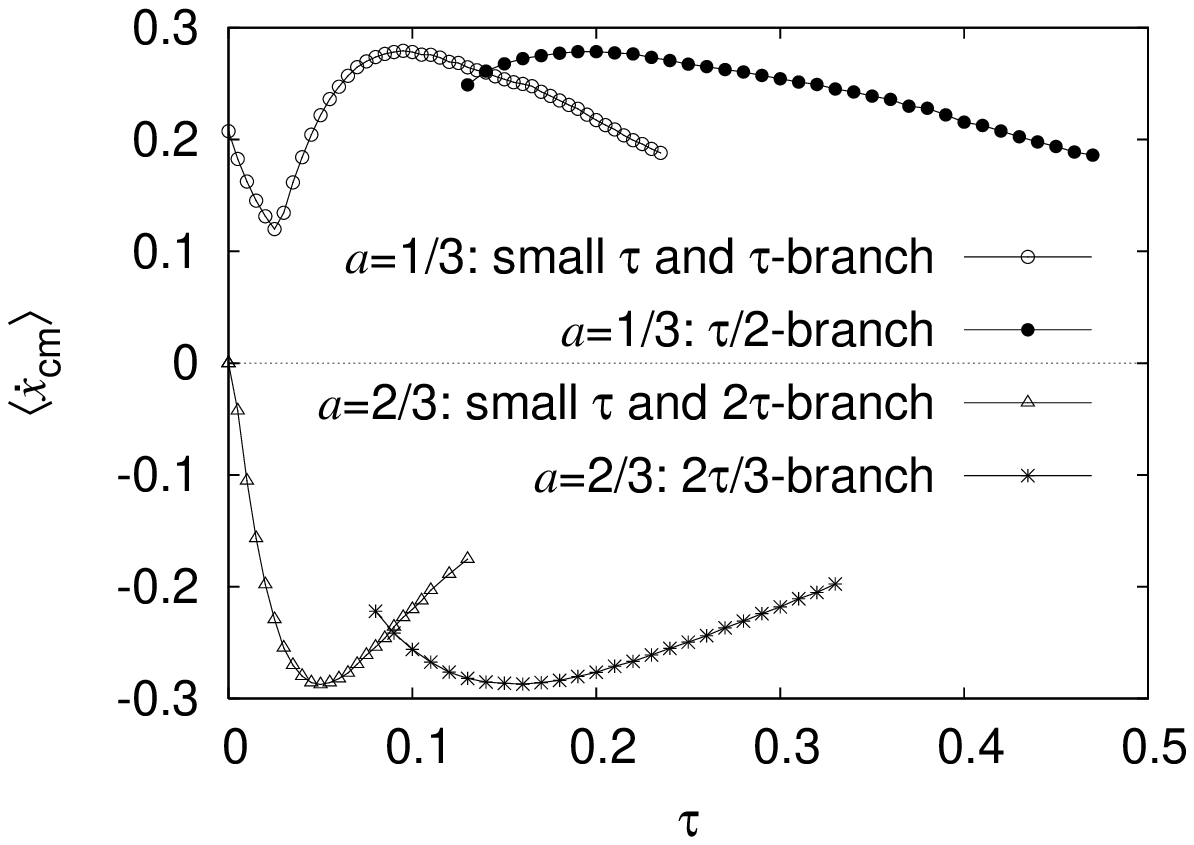}
\includegraphics[scale=0.5]{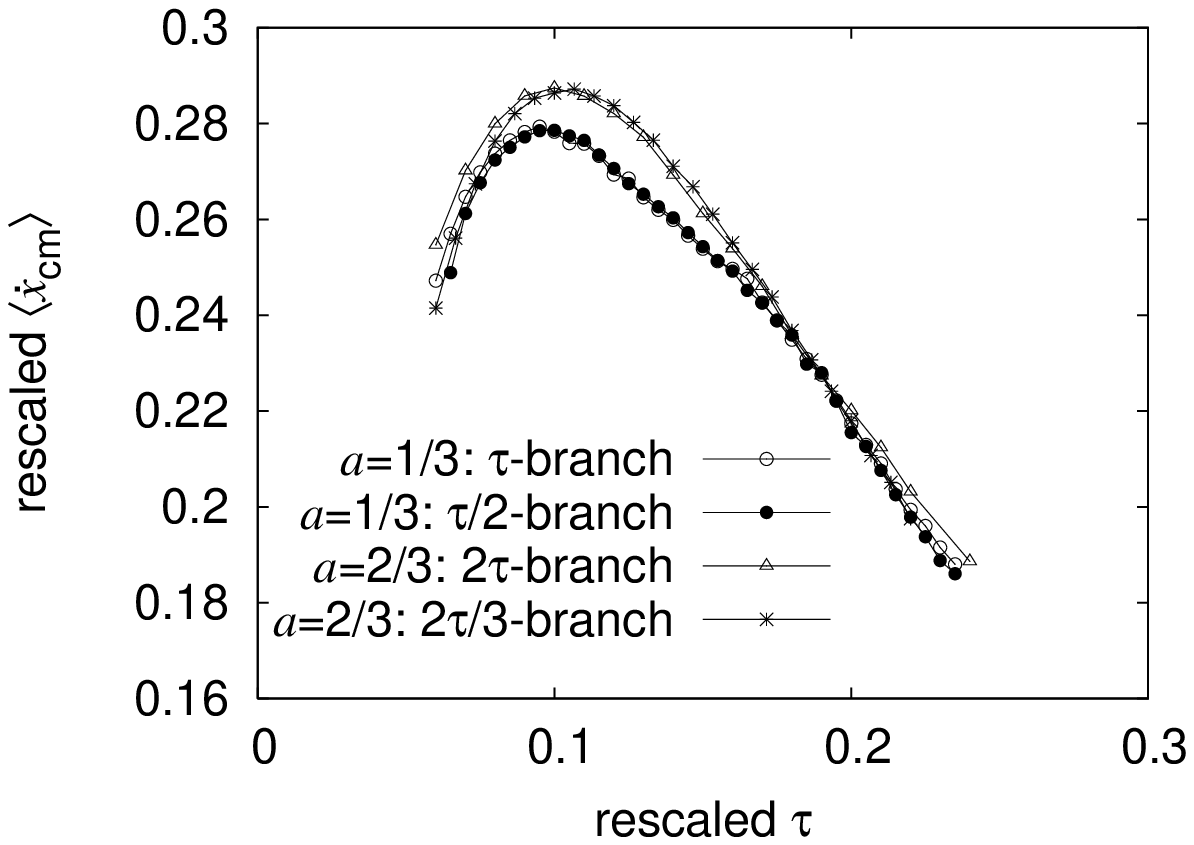}
\caption{\label{fig:branches}
Left panel: Center-of-mass velocity $\vmedia$ versus the delay $\tau$ in the many particle
case ($N=10^5$). The region of small delays and the first two
branches for asymmetries parameters $a=1/3$ (positive flux) and $a=2/3$
(negative flux) are plotted for height of the potential $V_0=5k_BT$. Units:
$L=1$ and $k_BT=1$.
Right panel: First two branches for asymmetries $a=1/3$ and $a=2/3$
for potential height $V_0=5k_BT$ and $N=10^5$ particles,
rescaled according to scaling laws~\eqref{relreversed0}
and~\eqref{relreversed}. 
}
\end{figure}

For larger delays there are stable solutions of quasiperiods
${\cal T}=2\tau/(2n+1),\;n=0,1,\dots$, i.e., solutions that
contain an odd number of semiperiods $ {\cal T}/2 $ in the time
delay $ \tau $. See Fig.~\ref{fig:evol} (right) for instance. The
competing asymmetry of the potential causes the stabilization of
those solutions where, due to the delay, the controller switches on
when the present net force is negative and switches off when it
would be positive, that is, the controller acts contrary to its
intentions and gives a negative flux. These branches are the
counterparts of the solutions of quasiperiods ${\cal
T}=\tau/n,\;n=1,2,\dots$ ($\tau$ containing an even number of
semi-quasiperiods) observed for asymmetries $a<1/2$~\cite{fei07b}.
The difference of one semiperiod is due to the effectively reversed
operation of the controller caused by the combined effect of the
competing asymmetry of the potential and the delay. Some of these
branches are plotted in Fig~\ref{fig:branches} for both the cases
of cooperation (positive currents) and competition (negative
currents) of asymmetries.

It is important to note that the average velocity for all these
branches can be reexpressed in terms of one of them. Let us define
$ g(\tau)\mathrel{\mathop :}=\vmedia_{\tau}(\tau)$ as the average velocity for
the branch of period ${\cal T}=\tau$, which is present for cooperative
potential asymmetry. For these asymmetries, $a<1/2$, we showed in
Ref.~\cite{fei07b} that the average velocities of the branches of periods 
${\cal T}=\tau/n$ are given by
\begin{equation}\label{relreversed0}
\vmedia_{\frac{\tau}{n}}(\tau) =
g\left(\tfrac{\tau}{n}\right)\quad \mbox{ for $a<1/2$}.
\end{equation}
On the contrary, for competing asymmetries, $a>1/2$, we have found here that
the solutions have quasiperiods ${\cal T}=2\tau/(2n+1)$, and furthermore, the
average velocities of these branches are given by
\begin{equation} \label{relreversed}
\vmedia_{\frac{2\tau}{2n+1}}(\tau) = -
g\left(\tfrac{2\tau}{2n+1}\right)\quad \mbox{ for
  $a>1/2$.}
\end{equation}
Consequently, given one of the branches all the others can be
predicted; see Fig.~\ref{fig:branches}. This also implies that the analytical
results obtained 
in Ref.~\cite{fei07b} for cooperative potential asymmetry ($a<1/2$) are
directly extended to the competing potential asymmetry case
($a>1/2$), just using the relation in Eq.~\eqref{relreversed}.

\section{Conclusions}\label{sec:discussion}

We have studied the performance of feedback flashing ratchets when there is
competition between the asymmetry in the potential and the asymmetry in the
control protocol, and we have also studied the effects of tuning their
relative influences in the dynamics by the introduction of a time delay. An
experimental realization of a flashing ratchet has been performed
in~\cite{rou94} by using polystyrene latex spheres of diameters $d\simeq
0.25-1\;\mu m$ in an aqueous solution [viscosity $\eta\simeq
10^{-3}\;  Pa\cdot s$; $D= k_BT/(3\pi\eta d)$] exposed to a sawtooth dielectric
potential of period $L\simeq 50\;\mu m$. This experimental setup can be
modified to become an experimental realization of a feedback flashing ratchet
by monitoring the particles with a conventional charge-coupled device (CCD)
of about $30 \mbox{ fps}$ and processing the images to switch on or off
the ratchet potential in accordance with the particle positions. The time
delays considered here are introduced by delaying the action of the controller
a time between $ \tau = 0.01 L^2/D \sim 10 s $ and $ \tau = 0.5 L^2/D \sim 500
s $. (For a more detailed discussion see Ref.~\cite{fei07b}.)
Indeed a sophisticated feedback control has been recently implemented
in Ref.~\cite{coh06}, where images of a Brownian particle are
acquired on a high-sensitivity CCD of up to 300 fps
and thereafter a software processes the information to extract the position of
the particle and apply a feedback voltage.
On the other hand, we highlight that the viscous friction coefficient $\gamma$
depends on the shape and the size of the Brownian particle. Thus, as  
the adimensional delay must be multiplied by the factor $L^2/D=\gamma
L^2/k_BT$ in order to recover physical units, Brownian particles 
of different shape and size respond differently to a given time delay.
This effect could be useful for separating different kinds of macromolecules.
\par

We have seen that the performance of the system with competing asymmetries
differs significatively from its counterpart ratchet with cooperating 
asymmetries. In the absence of delay the competition of
asymmetries implies a  decay of the current with the size of the ensemble
much stronger than in the cooperative case ($1/\sqrt N \text{ vs } 1/\ln
N$). In the presence of delay the dynamics becomes richer with a current
reversal for large delays. In the few particle regime the change 
from positive to negative current 
can be understood as a change from a purely closed-loop control to an 
effective open-loop control. On the other hand, in the many
particle case the negative current regime appears for large enough delays as
the consequence of the stabilization of several branches of quasiperiodic
solutions. These stable branches have the opposite sign and are one 
semiperiod displaced with respect to those obtained for cooperating 
asymmetries, they also have a direct relation with them that allows the
extension for the competing asymmetries case of the analytical results found
in Ref.~\cite{fei07b} for cooperative asymmetries.

\begin{acknowledgments}
We acknowledge financial support from the Ministerio de Ciencia y
Tecnolog\'{\i}a (Spain) through the Research Project
FIS2006-05895. In addition, M.F. thanks the Universidad
Complutense de Madrid (Spain) and F.J.C. thanks ESF Programme
STOCHDYN for their financial support.
\end{acknowledgments}


\end{document}